\documentclass[aps,twocolumn, floats,preprintnumbers,showpacs,prl]{revtex4}
\usepackage{graphicx}
\usepackage{url}
\usepackage{bm}
\usepackage{amsmath}
\usepackage{amsfonts}
\usepackage{amssymb}
\def\beq{\begin{equation}}
\def\eeq{\end{equation}}
\def\beqa{\begin{eqnarray}}
\def\eeqa{\end{eqnarray}}

\def\ltap{\ \raise.3ex\hbox{$<$\kern-.75em\lower1ex\hbox{$\sim$}}\ }
\def\gtap{\ \raise.3ex\hbox{$>$\kern-.75em\lower1ex\hbox{$\sim$}}\ }
\def\slash#1{#1\!\!\!/\!\,\,}  
\def\sla#1{#1\!\!\!\!\!/\,\,\,}

\newcommand{\order}{{\cal O}}

\begin{document}
\preprint{FERMILAB-PUB-10-001-T;
     EFI Preprint 10-9}

\title{Weakly Interacting Stable Pions}

\author{
 Yang Bai$^a$
 and Richard J. Hill$^{b}$ 
\\
\vspace{2mm}
${}^{a}$Theoretical Physics Department, Fermilab, Batavia, Illinois 60510 \\
${}^{b}$Enrico Fermi Institute and Department of Physics, The University of Chicago, Chicago, Illinois 60637
}



\begin{abstract}
  An unbroken discrete symmetry, analogous to $G$-parity in QCD,
  exists in standard model extensions with vector-like coupling of electroweak $SU(2)$
  to ``hidden sector'' fermions that
  are confined by a strong gauge force.  For an arbitrary irreducible $SU(2)$
  representation of the hidden sector fermions, the lightest hidden sector states form an
  isotriplet of ``pions'' with calculable mass splittings and couplings
  to standard model fields.  The parity can be extended to
  fermions in real representations of color $SU(3)$, and can provide
  dark matter candidates with distinct collider signatures.
\end{abstract}
\maketitle
%

\noindent
{\it{\textbf{Introduction.}}}
Cosmological and astrophysical observations indicate the existence of
dark matter.  This motivates the investigation of standard model (SM)
extensions with new stable particles.  For example, in theories with
an exact parity symmetry, the lightest parity-odd particle is stable
against decay, and can act as dark matter.  Here we describe the
implications of a parity selection rule that naturally arises in a
``hidden sector'' involving strongly-coupled fermions, e.g. a QCD-like
theory with large confinement scale.

At the renormalizable level, such a hidden sector can communicate with
the standard model only through gauge interactions: $SU(3)_c$,
$SU(2)_{W}$ or $U(1)_Y$.  To avoid spontaneous breaking of these
symmetries at a high scale we focus on vector-like gauging, i.e.,
identical coupling of the gauge fields to left- and right-handed
fermions.

We will see that the class of models with gauge-confined fermions
coupled vectorially to $SU(2)_W$ has some remarkably simple universal
features.  A parity symmetry acting only on the hidden sector fields
is left unbroken.
The new parity symmetry is an analog of $G$ parity in
QCD~\cite{G parity}, which is broken in the SM by the presence of
$U(1)_Y$ and non-vectorial coupling of $SU(2)_W$.  Below the hidden
sector confinement scale, the lightest physical states are
Nambu-Goldstone bosons (NGBs) of the spontaneously broken chiral
symmetries.  In the limit of vanishing electroweak coupling ($g_2\to
0$) the pions are degenerate in mass, as implied by exact global
symmetries.  At finite coupling, and after electroweak symmetry
breaking, radiative corrections induce mass splittings, and the
electric-neutral component of an $SU(2)_W$ triplet is the ``lightest
$G$-odd particle'' (LGP).  If such a hidden sector exists, this
particle can be a dark matter candidate.  Its mass and standard model
couplings are calculable in terms of a free parameter representing the
confinement scale of the hidden sector fermions.

Models with vector-like coupling of standard model gauge fields to
hidden sector matter have previously been considered from the point of
view of dark matter~\cite{Cirelli:2005uq} and
collider~\cite{Kilic:2009mi} phenomenology.  In a large class of
models, our conclusions differ from previous work.  We demonstrate how
stable particles naturally arise in such models without ad hoc
assumptions, and provide a chiral lagrangian framework for
systematically computing properties of the new particles.

\noindent
{\it{\textbf{Fermion description.}}}
Let us extend the standard model by introducing Dirac fermions in a
complex representation of a hidden sector gauge group, e.g. the fundamental 
of $SU(N)_h$ with $N\ge 3$,
i.e., a QCD-like theory.  Suppose there are $n_f=2j+1$ flavors
transforming in an irreducible isospin $j$ representation of $SU(2)_W$~\footnote{
We assume that $N$ and $n_f$ are such that spontaneous chiral symmetry breaking
occurs.  We are not here concerned with issues such as perturbativity and 
unification at high scales.   
}.
The basic lagrangian is
\begin{align} {\cal L} = {\cal L}_{\rm SM} -\frac14
  (\hat{F}^a_{\mu\nu})^2\! + \! 
  \bar{\hat{\psi}}\left(\! i\slash{\partial}
    \!+\! \hat{g}\slash{\hat{A}}^b \hat{t}^{b} \! +\! g_2 \sla{W}^a\! J^a
  \right)\hat{\psi} \,,
  \label{eq:Lagrangian2flavor}
\end{align}
where hatted quantities refer to the hidden sector.  The Hermitian
matrices $\hat{t}^b$, $b= 1\dots N^2-1$, denote generators in the
fundamental $SU(N)$ representation and $J^a$, $a= 1\dots 3$, are
$n_f\times n_f$ isospin matrices.

All representations of $SU(2)$ are real.  The matrix $S$ that relates
the original and conjugated generators, $S^\dagger J^a S = -J^{a*}$,
can be expressed in terms of a group element~\cite{group theory}:
$S=\exp[i\pi J^2]$.  On the basis of isospin $j$ states, $|j m\rangle$
with $m=j \dots -j$, we have
\begin{align}\label{eq:S}
  S_{m_1m_2} = (-1)^{j+m_2}\delta_{m_1, -m_2} = \left(
    {\rm{exp}}\,\left[i\,\pi\,J^2\right] \right)_{m_1 m_2} \,.
\end{align}
For example, for $j=1/2$ we have $J^a=\tau^a/2$ and $S=i\tau^2$ where
$\tau^a$ are the usual Pauli matrices~%
\footnote{
This example was discussed in Ref.~\cite{Kilic:2009mi}, where it was 
incorrectly stated that stable particles are absent for $j>1/2$. 
}.  We use the reality of $SU(2)$
representations to define a modified charge conjugation operation,
``$G$-parity'',
\begin{align}
  \label{eq:G}
  \hat{\psi}&\xrightarrow{G} S\, \hat{\psi}^{\cal C} = S\, i \gamma^2 \hat{\psi}^*  \,, \nonumber \\
  \hat{A}^b\,\hat{t}^b &\xrightarrow{G} (\hat{A}^b)^{\cal C}\,\hat{t}^b = \hat{A}^b\,(-\hat{t}^{b*}) \,,
\end{align}
with $\cal C$ denoting ``ordinary'' charge conjugation and $S$ acting
on the fermion flavor index.  All SM fields, in particular $W^a_\mu$,
are left invariant.  It is readily verified that the lagrangian
(\ref{eq:Lagrangian2flavor}) is invariant under the transformation
(\ref{eq:G})~\footnote{The path integral measure is similarly
  invariant, so that the parity is well defined at the quantum
  level.}.  $G$-parity is a good quantum number of the theory, and all
SM particles are $G$-even.

\noindent
{\it{\textbf{Pion description.}}}
The $SU(N)_h$ gauge theory is assumed to have a behavior similar to
QCD. The gauge coupling becomes strong in the infrared, triggering
confinement and chiral symmetry breaking at a scale $\Lambda_h$. Below
$\Lambda_h$, the effective theory is described by ``pions'' which are
NGBs associated with the spontaneously broken global flavor symmetry
of the hidden sector: $SU(2j+1)_L\times SU(2j+1)_R\times U(1)_{B} \to
SU(2j+1)_V\times U(1)_{B}$. Here $U(1)_{B}$ is the unbroken dark
baryon number symmetry.

The pions are collected into the $SU(2j+1)$ matrix field 
$U \equiv e^{2 i\tilde{\Pi}/f_\Pi}$, where 
$f_\Pi$ is the analog of $f_\pi \approx 93\,{\rm MeV}$ in QCD. 
The $(2j+1)^2-1=\sum^{2j}_{J=1}\,(2J +1)$ pions can be classified in 
irreducible representations, $J$, of $SU(2)_W$. 
The generators $t^{(JM)}$ associated with pions of definite total isospin $J$ 
and electric charge $M$ 
can be expressed using Clebsch-Gordan coefficients for $j\times j$ as
\begin{equation}
  \label{eq:JM} t^{(JM)}_{m_1 m_2} \,=\, (-1)^{j-m_2}\, \langle jj J
  M |jj m_1, -m_2 \rangle \,.  
\end{equation}
The generators (\ref{eq:JM}) are normalized as
$\mbox{Tr}[t^{(JM)}t^{(J^\prime M^\prime)}]=(-1)^M \delta^{J
  J^\prime}\delta^{M,-M^\prime}$, are real by definition, and satisfy
$[t^{(JM)}]^T = (-)^M t^{(J,-M)}$.  The canonically normalized isospin
generators appearing in (\ref{eq:Lagrangian2flavor}) are
$J^3=\sqrt{C(j)} t^{(10)}$, $J^\pm /\sqrt{2}= \mp \sqrt{C(j)} t^{(1
  \pm 1)}$, where $J^\pm = J^1 \pm i J^2$, and $C(j)=j(j+1)(2j+1)/3$
is the normalization constant for the isospin-$j$ representation: ${\rm
  Tr}(J^a J^b)=C(j)\delta^{ab}$.  The pion field can be expanded as
\begin{equation} 
  \tilde{\Pi} = \sum_{i=1}^{(2j+1)^2-1} \Pi^i t^i = \sum_{J=1}^{2j}
  \sum_{M=-J}^{J} \Pi^{(JM)} t^{(JM)} \,.  
\end{equation}
Under the transformation (\ref{eq:G}) and using (\ref{eq:S}) we have
\begin{equation}
  \Pi^{(JM)} \xrightarrow{G} (-1)^J \Pi^{(JM)} \,,
\end{equation}
so that pions with odd (even) $J$ are odd (even) under $G$ parity. 
Note that $G$-parity generalizes the notion of pion number 
parity in QCD to arbitrary number of flavors.

\noindent
{\it{\textbf{Spectrum.}}}
The leading, two-derivative, term in the symmetric chiral Lagrangian is 
\begin{align} \label{pppp}
  {\cal L}_2 &=\frac{f^2_\Pi}{4}\,{\rm Tr} [ D_\mu U D^\mu U^\dagger  ] \,,
\end{align}
where $D_\mu U \equiv \partial_\mu U - i\,g_2\,W^a_\mu\,[J^a, U]$.  This
term describes massless, self-interacting scalar fields.  Note that
the neutral pions $\Pi^{(J 0)}$ correspond to generators commuting
with $J^3$, and hence have no direct interactions with the physical
photon or $Z$ boson.  The gauge coupling $g_2$ explicitly breaks the
global chiral symmetry.  At one loop, a quadratically divergent
counterterm is required,
\begin{align}\label{Lgexpand} 
  {\cal L_{\rm g}} \!&\sim \, g_2^2
  \Lambda_h^2 f_{\Pi}^2 \,{\rm Tr}( J^aU J^a U^\dagger ) \,.
\end{align}
This loop effect lifts the degeneracy among multiplets,
\begin{equation}
  \label{eq:pimass} m^2_{\Pi^{(JQ)}} \sim J(J+1) \alpha_2
  \Lambda_h^2 \,, 
\end{equation}
where $g_2^2\equiv 4\pi \alpha_2$.  The hidden
sector couples to the SM Higgs field via the $SU(2)_W$ gauge field
intermediary.  After electroweak symmetry breaking, such interactions
induce a finite splitting between components of each multiplet.  In
the limit $m_\Pi \gg m_W$, 
$m_{\Pi^{(J Q)}} - m_{\Pi^{(J 0)}} \approx \alpha_2  Q^2  m_W \sin^2{\theta_W \over 2}$~\cite{Cirelli:2005uq, baihill}.
In particular, $m_{\Pi^{(1 \pm 1)}} - m_{\Pi^{(1 0)}} \approx 170\,{\rm MeV}$, leaving 
$\Pi^{(1 0)}$ as the LGP.   

\noindent
{\it{\textbf{Interactions.}}}
The interactions of the pions amongst themselves, and with SM fields,
are constrained by chiral symmetries of the new strong interaction.
Taking over results familiar from QCD, the symmetric lagrangian can be
expanded in powers of $1/f_\Pi$: ${\cal L} = {\cal L}_2 + {\cal L}_4 +
\dots$.  The leading term is displayed in (\ref{pppp}).  Of particular
interest at four-derivative order is the Wess-Zumino-Witten (WZW)
interaction, containing terms with odd pion number,
\begin{multline}
  \label{WZW} {\cal L}_4 = \frac{N g_2^2 }{16\pi^2
    f_\Pi}\epsilon^{\mu\nu\rho\sigma} {\rm Tr} \left[ {\tilde{\Pi}} W_{\mu\nu}
    W_{\rho\sigma} + \order(\tilde{\Pi}^3) \right ] + \dots \,.
\end{multline}

Let us consider the decays of the lightest states, focusing for
simplicity on the limit $m_\Pi \gg m_W$.  The charged members of the
ground state multiplet decay with weak strength into the LGP via QCD
pion emission: 
\begin{equation}
  \Gamma(\Pi^{(1\pm 1)} \to \Pi^{(10)} + \pi^\pm) =
  {4G_F^2 \over \pi} \Delta^2\sqrt{\Delta^2-m_{\pi}^2} f_\pi^2 \,, 
\end{equation}
where $\Delta=m_{\Pi^{(1\pm 1)}} - m_{\Pi^{(1 0)}}$ and we have
assumed $m_\Pi \Delta \gg m_\pi^2$.  In this limit the decay length
$c\tau \approx 5\,{\rm cm}$ is independent of $m_\Pi$.  The
lightest $G$-even pions, $\Pi^{(2 M)}$, decay into electroweak bosons
via the WZW term (\ref{WZW}).  The relevant trace is
\begin{multline}
 C(j) {\rm Tr}\big[t^{(2 M)} \{ t^{(1 L)}, t^{(1 L^\prime)} \} \big] =
  (-)^{M} \langle 11 2 M | 11 -L^\prime -L\rangle
  \\
  \times {4\over \sqrt{15}}
  \left[ (j-1/2)j(j+1/2)(j+1)(j+3/2)\right]^{\frac12} \,.
\end{multline}
The amplitude is sensitive both to the number of colors $N$ and to the
isospin $j$ of the underlying fermions.  The rate scales as
$\Gamma(\Pi^{(2 M)}) \sim \alpha_2^2 m_\Pi^3 /(4\pi)^3 f_\Pi^2 $.
Note that the process $\Pi^{(1M)}\to WW$ (the analog of $\pi^0\to
\gamma\gamma$ in the SM) is forbidden by the vanishing of the relevant
isospin trace (``$d$ symbol'').

The next-to-lightest $G$-odd pions, $\Pi^{(3 M)}$, decay into the
ground-state multiplet via loops containing $G$-even pions, $\Pi^{(2
  M)}$.  The interaction vertices are obtained by expanding (\ref{pppp}) 
and (\ref{Lgexpand}). 
The decay rate scales as 
$\Gamma(\Pi^{(3 M )}\to \Pi^{(1 M^\prime)} + 2W ) \sim
\alpha_2^2 m_\Pi^5/ (4\pi)^5 f_\Pi^4$.  For $J\ge 3$, loops containing
$G$-even pions mediate the decays $\Pi^{(J M)}\rightarrow \Pi^{(J-2,
  M^\prime )} + 2W$.  The end result is a multi-$W$ final-state.  For
odd $J$, the stable $\Pi^{(1 0)}$ remains, possibly after decay of the
long-lived $\Pi^{(1\pm 1)}$ as above.

\noindent
{\it{\textbf{Coupling to $SU(3)_c$.}}}
The parity operation (\ref{eq:G}) can be extended to real
representations of $SU(3)_c$.  (Extension to $U(1)_Y$ would require
embedding the gauge fields inside a larger, real representation.)

Consider, e.g., hidden sector fermions transforming in the adjoint
representation of $SU(3)_c$, and the fundamental representation of
$SU(2)_W$.  In the usual basis where adjoint generators are purely
imaginary, the generalized $G$-parity for this model is defined as in
(\ref{eq:G}) where now $S= \openone_8 \otimes i \tau^2$ acting on
$SU(3)$ and $SU(2)$ indices.  Using the product decomposition
$\bm{8}\times \bm{8} = (\bm{8} + \bm{10} + \overline{\bm{10}})_{\rm A}
+ (\bm{1} + \bm{8} + \bm{27})_{\rm S}$ (``A" and ``S" denoting
antisymmetric and symmetric tensors) and $\bm{2}\times \bm{2}
=\bm{1} + \bm{3}$, we can express the $16^2-1=255$ pion generators as
products of $SU(3)$ and $SU(2)$ generators.  The resulting symmetric
(anti-symmetric) pure $SU(3)$ generators are $G$-even ($G$-odd) and the
decomposition into representations of $SU(3)_c\times SU(2)_W$ with the
corresponding $G$-parity is
$(\bm{8},\bm{2})\times(\bm{8},\bm{2})-(\bm{1},\bm{1}) = (\bm{1},\bm{3})^- +
(\bm{8}_{\rm A},\bm{1})^- + (\bm{8}_{\rm S},\bm{1})^+ + (\bm{8}_{\rm
  A},\bm{3})^+ + (\bm{8}_{\rm S},\bm{3})^- + \dots$.
The pion masses are proportional to the sum of second-order Casimir
invariants as in (\ref{eq:pimass}), $m^2_{\Pi^{(d_3, d_2)}} \sim \Lambda_h^2 \left[ \alpha_3 C_2(d_3)
    + \alpha_2 C_2(d_2) \right]$, 
with $C_2(3)=2$ for $SU(2)$, and $C_2(8)=3$ for $SU(3)$. The
lightest states, $(\bm{1},\bm{3})^-$, are again an $SU(2)_W$ triplet of $G$-odd pions.

Operators such as (\ref{pppp}), (\ref{Lgexpand}) and (\ref{WZW}) again
mediate transitions between pion states, after appropriate
substitution to account for $SU(3)_c\times SU(2)_W$ gauging.  The
decays of the two lightest $G$-even multiplets $(\bm{8}_{\rm S},\bm{1})^+$ and
$(\bm{8}_{\rm A},\bm{3})^+$, proceed primarily through WZW interactions.  The
two next-to-lightest $G$-odd pions, $(\bm{8}_{\rm A},\bm{1})^-$ and $(\bm{8}_{\rm S},\bm{3})^-$, 
decay to the LGP plus two SM gauge
bosons through loop diagrams with $G$-even pions.  The spectrum and
interactions in this model can produce interesting collider
signatures. For example, the pair production of the next to lightest
$G$-odd particle $(\bm{8}_{\rm A},\bm{1})^-$ leads to a final state
with 2 photons + 2 jets + missing transverse energy~\cite{baihill}, 
with displaced vertices
from intermediate metastable electrically-charged $(\bm{1}, \bm{3})^-$ states.

Other possibilities for the hidden sector include reducible
representations of $SU(2)_W$, i.e., multiple ``generations''.  These cases might be interesting to investigate from
the standpoint of SM gauge-singlet dark matter (interacting via
higher-dimension operators), or as nearly-degenerate iso-triplets.
Another possibility (not relevant to weakly coupled dark matter) is to
consider $SU(2)_W$ singlet fermions in an irreducible real
representation of $SU(3)_c$.
Various models can be embedded in five dimensional gauge theory constructions. 

\noindent
{\it{\textbf{Peccei Quinn symmetry and axion.}}}
We have so far neglected two gauge invariant terms in the lagrangian
(\ref{eq:Lagrangian2flavor}) that can appear at the renormalizable
level.  The first is a theta term for the hidden gauge fields, and the
second a bare mass term for the fermions~\footnote{ We have used the
  freedom to perform a chiral rotation of the fermions and redefine
  $\theta$ to eliminate any $\bar{\hat{\psi}}\gamma_5\hat{\psi}$ term.
},
\begin{equation}
  \label{eq:mtheta} {\cal L}_\theta = \theta
  \epsilon^{\mu\nu\rho\sigma} \hat{F}^a_{\mu\nu} \hat{F}^a_{\rho\sigma}
  \,, \quad {\cal L}_m = - m_\psi \bar{\hat{\psi}}\hat{\psi} \,.  
\end{equation}
For nonzero $m_\psi$ and $\theta$, these terms give rise to $P$ and
$T$ violation in the hidden sector but do not affect $G$-parity
conservation~\footnote{ A Majorana mass term violating $G$ parity is
  forbidden by gauge invariance for complex representations of
  $SU(N)_h$.  }.  For simplicity we focus on the case $m_\psi \ll
\Lambda_h$.  For $m_\psi \gg \Lambda_h$, the lightest $G$-odd states
become iso-triplet nonrelativistic ``quarkonium''.

It is natural to consider whether some mechanism suppresses $P$ and
$T$ violation in the hidden sector, as happens in QCD.  Note that at
$m_\psi=0$, there is a Peccei-Quinn (PQ) symmetry present at the
classical level, $\hat{\psi} \to e^{i\alpha\gamma_5} \hat{\psi}$. The PQ symmetry can be preserved at $m_\psi \ne 0$  if the mass is generated by spontaneous symmetry breaking. Consider a scalar field $\sigma$ that transforms under the PQ symmetry as 
$\sigma\to e^{-2i\alpha}\sigma$, and with interactions:
\begin{equation}
  {\cal L}_\sigma = |\partial_\mu \sigma|^2 -V(\sigma) -\lambda
  \sigma \bar{\hat{\psi}}_L \hat{\psi}_R + {\rm h.c.} \,,
\end{equation}
where $V(\sigma)$ is such that $\sigma$ acquires a VEV,
$\langle\sigma\rangle = f_a$.  The mass parameter is then $m_\psi=
\lambda f_a$, and with $\sigma(x) \sim f_a e^{i a(x)/\sqrt{2} f_a}$,
the low-energy spectrum includes an axion $a(x)$ which is massless at
the classical level.

As with the QCD axion, we assume that the $SU(N)_h$ anomaly generates
an effective potential for $a(x)$ with minimum such that the effective
$\theta$ term vanishes.  Gauge invariance prevents the hidden axion
from direct interaction with SM fermions~\footnote{ The PQ symmetry
  could be extended to accommodate a right-handed neutrino coupling by
  assigning a common vector PQ charge to all SM leptons.  }.  The
physical axion acquires mass and interactions
through mixing with the NGB of the $U(1)_A$
global flavor symmetry in the hidden sector,
\begin{equation} {\cal L}_a = {m_a^2 \over 2 f_\Pi^2} [\Pi^{i}]^2
  a^2 - \frac{N C(j)
    g_2^2}{32\pi^2\sqrt{2}f_a} a \epsilon^{\mu\nu\rho\sigma} W^a_{\mu\nu}
  W^a_{\rho\sigma} + \dots \,,
\end{equation}
where $m_a^2 \sim m_\psi \Lambda_h f_\Pi^2/f_a^2$.  Note that the
$SU(2)_W$ couplings to the hidden sector do not break the $U(1)_{PQ}$
symmetry, and so do not contribute to the axion mass: only in the
limit $\sqrt{m_\psi\Lambda_h} \gg g_2 \Lambda_h$ does the QCD-like
relation $m_a f_a \sim m_\Pi f_\Pi$ hold.

\noindent
{\it{\textbf{Discussion.}}}
Hidden sector fermions coupled vectorially to SM gauge fields are an
ingredient in typical axion models addressing the strong CP
problem~\cite{Kim:1979if}.  It is natural to inquire whether other
such hidden sectors exist.  We have considered approximately massless
Dirac fermions in a complex representation of a QCD-like gauge group,
and an arbitrary vector-like coupling to $SU(2)_W$.  The lightest
state of such a hidden sector is the neutral component of an
approximately degenerate iso-triplet ``pion''. In contrast to the $C$ and 
$CP$ symmetries in the SM, there are no gauge invariant renormalizable 
operators that violate the new ``$G$" parity.

Could weakly interacting stable pions (WISPs) be a component of
cosmological dark matter, produced thermally in the early universe?  A
lifetime of order $\tau_{\rm universe} \sim 10^{10} \,{\rm years}$
imposes tight constraints and we must consider whether $G$ parity
could be broken either explicitly (by unknown UV physics) or
spontaneously (from the choice of vacuum).
Suppose that the SM extension (\ref{eq:Lagrangian2flavor}) is viewed
as an effective theory valid up to some scale $\Lambda_{\rm UV}$.
Corrections to the renormalizable lagrangian can appear at dimension
five.  Two such operators,
$B_{\mu\nu}\bar{\hat{\psi}}\sigma^{\mu\nu}\hat{\psi}/\Lambda_{\rm UV}$
and $H^\dagger t^a H \bar{\hat{\psi}} J^a \hat{\psi}/\Lambda_{\rm
  UV}$, violate $G$-parity.  Even for $\Lambda_{\rm UV}$ of order the
Planck scale, an additional suppression would be
necessary to ensure cosmological stability.  Note however that these
operators violate the PQ symmetry present at the renormalizable level.  
Enforcing this
symmetry implies the appearance of the $\sigma$ field, leading to an
additional suppression $\langle \sigma(x) \rangle /\Lambda_{\rm UV} \sim
f_a/\Lambda_{\rm UV}$.  Then $\tau_{\rm LGP} \sim \Lambda_{\rm UV}^4/ f_\Pi^3
f_a^2 \sim \tau_{\rm universe} \times 10^8[f_\Pi/f_a]^2$, for
$\Lambda_{\rm UV}\sim 10^{16}\,{\rm GeV}$ and $f_\Pi \sim 10^3 \,{\rm
  GeV}$.  Provided that $f_a/f_\Pi$ is not too large, the lifetime of
the LGP is plausibly long enough to be a thermal dark matter candidate.

Consider the vacuum alignment question: can the fermion condensate
$\Sigma^{ij}_0 \sim \langle \bar{\hat{\psi}}^i_L \hat{\psi}^j_R
\rangle$ spontaneously break $G$ parity?  For massless fermions, the
effective potential that determines $\Sigma_0$ begins at order
$g_2^2$.  As indicated by the positive pion masses (\ref{eq:pimass}),
the potential has a minimum at $\Sigma_0 \propto \openone$.  This conclusion
is unaffected if a bare fermion mass (\ref{eq:mtheta}) is included in
the hidden sector: the gauged vector-like symmetries are
unbroken~\cite{Vafa:1983tf}.  Interactions with the SM lead to
$\order(g_2^4)$ corrections to the effective potential.  Since $G$
parity is not explicitly broken, the potential is invariant under $G$
parity, and by a simple continuity theorem~\cite{Georgi:1974au}, the
shift (if any) of $\Sigma_0$ is similarly invariant.  Barring a
conspiracy of higher-order terms, $G$ parity should not be
spontaneously broken.

What about hidden sector baryons?  These are also stable particles
and potentially contribute to the dark matter relic abundance.   
In the absence of hidden baryon nonconservation, 
baryon-antibaryon pairs are created in thermal equilibrium.   
At freezeout, the energy density 
of baryons to pions scales as~\cite{baihill} $\Omega_B/\Omega_\Pi \sim 
\langle \sigma v \rangle_{\Pi} / \langle \sigma v\rangle_B \sim 4\pi \alpha_2^2 \sim 10^{-2}$, 
indicating that the baryons play a subdominant, but non-negligible role. 
 (Here $\sigma$ denotes annihilation cross section).   
In cases where $N$ and $j$ forbid a neutral baryon, however, the stable 
baryons can be fractionally charged, imposing tight constraints.  Models of (techni-)baryonic dark matter have been investigated in~\cite{Frandsen:2009mi}.  

What are the existing constraints on WISPs?  The impact of a
vector-like hidden sector on low-energy precision measurements is
minimal~\cite{Lavoura:1992np,Kilic:2009mi}.  A search for charged fermions decaying to missing energy at
LEP constrains $m_{\rm LGP} \gtrsim 88\,{\rm GeV}$~\cite{Heister:2002mn}.
Studies of nearly degenerate fermion isotriplets suggest a potential
sensitivity up to $\sim 100\,{\rm GeV}$ at the
Tevatron~\cite{Feng:1999fu}, and somewhat higher mass at the
LHC~\cite{Buckley:2009kv}; production channels involving $SU(3)_c$
octets would increase the reach.  The precise values of the bounds
will depend on production cross sections for isotriplet scalars versus
fermions.  Studies of $SU(2)_W$ multiplets of simple scalar dark
matter particles~\cite{Cirelli:2005uq} suggest a bound $m_{\rm LGP} \lesssim
1-2\,{\rm TeV}$ in order to not overclose the universe, and a
spin-independent scattering cross section on a nucleon $\sim
10^{-45}~\mbox{cm}^2$; such particles should have so far evaded direct
dark matter detection experiments~\cite{Ahmed:2009zw}.  In all of these constraints,
precise values are affected by pion self-interactions, and possibly by
axion or color $SU(3)_c$ interactions.  We conclude that a
large window $\sim 100\,{\rm GeV} - {\rm few\, TeV}$ potentially exists for the LGP
mass.  A more thorough analysis will be presented elsewhere.

What are the constraints on the hidden sector axion?  The lifetime,
$\tau_a \sim [\alpha^2 m_a^3/f_a^2]^{-1}$, should be such that axions
produced thermally in the early universe decay before big bang
nucleosynthesis.  This avoids constraints from elemental abundances,
CMB distortions and diffuse photon backgrounds~\cite{Masso:1997ru}.
Assuming for illustration $m_\psi \sim m_\Pi \sim 100\,{\rm GeV}$ this
yields $f_a \lesssim 10^{4-5} f_\Pi$.  Note that if the axion mass is
not small compared to $g_2 f_\Pi$, the LGP can annihilate into two
axions with substantial branching fraction, influencing the thermal
history and potential dark matter annihilation signals.

In conclusion, we have argued that an unbroken and nontrivial
discrete symmetry emerges in a class of models with vector-like
SM gauging of fermions with QCD-like strong dynamics.  The mechanism leads to
testable and distinct dark matter and LHC phenomenologies.
 
\acknowledgments 
 We thank W.~Bardeen, M.~Carena, P.~Fox, A.~Martin and especially C.~Hill
for discussions.  
Research supported by NSF Grant No. 0855039 and DOE
contract no. DE-AC02-07CH11359.


\end{document}